\documentclass[intlimits,twoside,a4paper]{article}

\usepackage[utf8]{inputenc}
\usepackage[eqsecnum]{cmpj3}

\usepackage{bm}
\issue{2023}{26}{2}{23706}
\doinumber{10.5488/CMP.26.23706}

\title[Electronic structure and X-ray magnetic circular
dichroism]{Electronic structure and X-ray magnetic circular dichroism in
  the MAX phases T$_2$AlC (T=Ti and Cr) from first principles}

\author[L. V. Bekenov, S. V. Moklyak,
B. F. Zhuravlev, Yu. N. Kucherenko,
V. N. Antonov]{L. V. Bekenov\orcid{0000-0003-1839-0134}, S. V. Moklyak,
  B. F. Zhuravlev, Yu. N. Kucherenko,
  V. N. Antonov\orcid{0000-0003-3410-1582}}

\address{ G. V. Kurdyumov Institute for Metal Physics of the
  NAS of Ukraine, 36 Academician Vernadsky Boulevard, UA-03142 Kyiv,
  Ukraine}

\newcommand{\dxy}{\ensuremath{d_{xy}}}
\newcommand{\dxz}{\ensuremath{d_{xz}}}
\newcommand{\dyz}{\ensuremath{d_{yz}}}

\newcommand{\dxxyy}{\ensuremath{d_{x^2-y^2}}}

\newcommand{\mb}{\ensuremath{\mu_{\text{B}}}}

\date{Received December 20, 2022}

\begin{document}

\maketitle

\begin{abstract}

We study the electronic and magnetic properties of T$_2$AlC (T=Ti and
Cr) compounds in the density-functional theory using the generalized
gradient approximation (GGA) with consideration of strong Coulomb
correlations (GGA+$U$) in the framework of the fully relativistic
spin-polarized Dirac linear muffin-tin orbital (LMTO) band-structure
method. The X-ray absorption spectra and X-ray magnetic circular
dichroism (XMCD) at the Cr $L_{2,3}$ and Cr, Ti, and C $K$ edges were
investigated theoretically. The calculated results are in good
agreement with experimental data. The effect of the electric
quadrupole $E_2$ and magnetic dipole $M_1$ transitions at the Cr $K$
edge has been investigated.



\end{abstract}

\section{Introduction}

\label{sec:introd}

The M$_{n+1}$AX$_n$ (MAX) phases are layered hexagonal compounds, in
which close-packed layers of M (early transition metals) are
interleaved with layers of group A element (mostly IIIA and IVA), with
the X-atoms (C and/or N) filling the octahedral sites between the M
layers. Depending on the stoichiometry, one gets three different kinds
of crystal structures classified as 211 ($n = 1$), 312 ($n = 2$) and 413
($n = 3$), all being described by the $P6_3/mmc$ space group~\cite{IFR16}. These carbides and nitrides have received increasing
interest in recent years due to their important properties, such as
low density, high elastic stiffness, thermal-shock and
high-temperature oxidation resistance, good thermal and electrical
conductivities, damage tolerance, and easy machinability
\cite{Bar00,BaEl01,WaZh09,SAL+09,book:Barsoum10,EBJ+10}. These unusual
properties make these compounds highly promising candidates for
diverse applications in high-temperature oxidizing environments.
Potential applications of the nanolaminated compounds mentioned to
date range from spintronics to refrigeration, even though the research
efforts have so far been focused solely on the discovery of new
magnetic phases and compositions, and fundamentals of magnetic
properties.  However, individual properties vary from phase to phase
and no systematic study of MAX phases is available in the literature.
Up till now, more than 70~MAX compounds have been synthesized
\cite{MRC12,LMC13}, and the search is still going on. Moreover, an
increasing number of scientists attempt to synthesize magnetic MAX
phases \cite{DAA+11} because magnetic MAX phases exhibit a combination
of magnetic, metallic, and ceramic-like properties
\cite{DAR13,MDE+13}, which are fascinating from the standpoints of
both fundamental science and technological applications.  For example,
a substantial progress could happen in magnetic multilayers for
magnetic recording and data storage if strong magnetism can be found
in MAX phases with inherently nanolaminated compounds instead of the
presently used artificially grown sandwich materials.

The ternary carbides have the properties of both metals and ceramics. Like metals, they are good thermal and electrical
conductors. They are relatively soft with vickers hardness of about
2--5~GPa. Like ceramics, they are elastically stiff. Some of them
(e.g., Ti$_3$SiC$_2$) also exhibit excellent high temperature
mechanical properties. They are resistant to thermal shock and show
unusually good damage tolerance and exhibit excellent corrosion
resistance. Above all, unlike the traditional carbides, they can be
machined by conventional tools without lubricant, which makes them
more technologically important for applications~\cite{book:Barsoum10,SLA12}. These excellent properties make the MAX
phases another class of technically important materials. Brushes in
electric motors is another application of these materials. The
enormous applications of this class of materials trigger us to have a
better understanding of their electronic and mechanical properties.

Of special interest to this work are Ti$_2$AlC and Cr$_2$AlC carbides. The
Ti$_2$AlC phase was first synthesized by Nowotny with co-workers \cite{JNB64},
but its physical properties were not characterized until quite recently~\cite{Bar00,JNB64,SSL71,SEB02}. The electronic structure of Ti$_2$AlC was
calculated a number of times \cite{Mat97,HuFri02,ZhSu00}. The qualitative
agreement between the first and last studies is good. The ternary
nanolaminated compound Cr$_2$AlC was discovered in 1963 by Jeitschko et~al.~\cite{JNB63}. Since that time it has been the subject of numerous
studies, both experimentally
\cite{Now71,SSM+04,HLF+05,MGS+06,LLX06,LHM+06,LZZ+06,SSM+07,LHL+10,ABT+11,AZG+08,JBJ+14}
and theoretically
\cite{LMC13,DAA+11,DAR13,MDE+13,SLA12,SSM+04,SAL+03,LHH+04,WaZh04,SMA+04,HJB05,MuSc07,Bou09,ZLW09,JiYa10,DSH+11,RLA11}.

In spite of many experimental and theoretical efforts to investigate the electronic structure and physical properties of the ternary MAX phases, there are still many contradictions and discrepancies among them. Many questions still remain with respect to magnetism in the the ternary nanolaminated compounds (see the review paper by Ingason et~al.~\cite{IFR16} and references therein). Schneider et~al.~\cite{SSM+04} calculated the difference in cohesive energy per
formula unit (f.u.) of the ferromagnetic (FM) and antiferromagnetic
(AFM) configurations with reference to the nonmagnetic (NM)
configuration. Based on the small total energy difference between the
AFM and NM configurations as well as the comparatively small local
magnetic moment (0.7~{\mb}/Cr~atom), the authors speculated that magnetism
may be suppressed in Cr$_2$AlC. Sun et~al.~\cite{SAL+03} also
performed total energy calculations for Cr$_2$AlC both at AFM and NM
polarized states. They found that energy difference between the two
states was around 0.3~meV. Therefore, they claimed nonmagnetic ground
state in Cr$_2$AlC. This is in agreement with the first experimental
result attempting to address this point based on nuclear magnetic
resonance (NMR) spectroscopy~\cite{LLX06}. In this study, \mbox{Lue et~al.} show, from the temperature dependence of the $^{17}$Al central
lines between room temperature and 77~K, that Cr$_2$AlC is NM in this
temperature range. Ramzan et~al. \cite{RLA11} presented the
electronic and structural properties calculated by first principles
GGA and GGA+$U$ calculations of Cr$_2$AlC. They show that for the GGA
exchange-correlation potential, the ground state of Cr$_2$AlC is not spin-polarized, with vanishing magnetic moments on the Cr atoms. In this case, the agreement with experiments for the equilibrium
structure and the bulk modulus is not satisfactory. On the other hand, using the GGA+$U$ approximation (with Hubbard $U = 1.95$~eV and exchange Hund coupling $J = 0.95$~eV), the calculated ground state
corresponds to a FM ordering. In this case, the agreement with
experiments was found to be excellent for the lattice constants and bulk modules. However, for a larger value of $U$ (2.95~eV), the ground
state magnetic order is modified, from FM (for $U = 1.95$~eV) to AFM
(for $U = 2.95$~eV), and the calculated equilibrium volume was in
strong disagreement with experiments. The magnitude of the magnetic
moments strongly depends on the Hubbard $U$. Dahlqvist et~al.
\cite{DAR13,DAR15} state that the magnetic ground state of Cr$_2$AlC
is in-plane AF. By comparing GGA and GGA+$U$ results with experimental
data, they found that using a $U_{\rm eff}=U-J$ value larger than 1~eV
results in structural parameters deviating strongly from
experimentally observed values. The spin magnetic moments at the Cr
site $M_s^{\rm Cr}$ were found to be equal to 0.7~{\mb}~\cite{DAR13} and
1.37~eV for the GGA and GGA+$U$ (for $U_{\rm eff}= 1$~eV) \cite{DAR15},
respectively. The authors concluded that this class of Cr-based carbide
MAX phases cannot be considered as strongly correlated systems since
both GGA and GGA+$U$ with $U_{\rm eff} \leqslant 1$~eV give the calculated lattice
parameters and bulk modulus close to experimentally reported values,
if low-energy in-plane AFM magnetic states are considered. For larger
values of the $U$ parameter ($U_{\rm eff} > 1$~eV), the structural
parameters deviate strongly from experimentally observed
values.

Soft X-ray absorption and magnetic circular dichroism in T$_2$AlC were measured by several authors~\cite{HJB05,LZZ+06,JBJ+14}. Recent X-ray magnetic circular dichroism experiments performed by Jaouen et~al.~\cite{JBJ+14} demonstrate that Cr atoms carry a net magnetic
moment in Cr$_2$AlC ternary phase along the $c$ axis. The spin
magnetic moment at the Cr site was found to be extremely small
$M_s^{\rm Cr}=0.05$~{\mb}, which is much smaller than the predicted values for
the FM \cite{SSM+04,RLA11} as well as the AFM \cite{DAR13,DAR15}
solutions. SQUID experiments by Jaouen et~al. \cite{JCC+13} also
produce an extremely small Cr spin magnetic moment of 0.002~{\mb} in
Cr$_2$AlC.

The aim of this paper is the theoretical study from the ``first
principles'' of the electronic and magnetic structures and the XAS and
XMCD in T$_2$AlC (T=Ti and Cr) carbides. The energy band structure of
T$_2$AlC (T=Ti and Cr) carbides is calculated within the {\it ab
  initio} approach taking into account strong electron correlations by
applying a local spin-density approximation to the density functional
theory supplemented by a Hubbard $U$ term (GGA+$U$) \cite{AZA91}. The
paper is organized as follows. The computational details are presented
in section~\ref{sec:details}. Section~\ref{sec_3} presents the electronic structure, XAS and XMCD
spectra of T$_2$AlC (T=Ti and Cr) carbides at the Cr $L_{2,3}$ and Cr,
Ti, and C $K$ edges calculated in the GGA+$U$
approximation. Theoretical results are compared to the experimental
measurements. Finally, the results are summarized in section~\ref{sec:summ}.

\section{Crystal structure and computational details}
\label{sec:details}

\subsection{X-ray magnetic circular dichroism} 

Magneto-optical (MO) effects refer to various changes in the
polarization state of light upon interaction with materials possessing
a net magnetic moment, including rotation of the plane of linearly
polarized light (Faraday, Kerr rotation), and the complementary
differential absorption of the left-hand and right-hand circularly polarized light
(circular dichroism). In the near visible spectral range, these effects
result from excitation of electrons in the conduction band. Near X-ray
absorption edges, or resonances, magneto-optical effects can be
enhanced by transitions from well-defined atomic core levels to
transition symmetry selected valence states.

Within the one-particle approximation, the absorption coefficient
$\mu^{\lambda}_j (\omega)$ for incident X-ray polarization~$\lambda$ and
photon energy $\hbar \omega$ can be determined as the probability of
electronic transitions from initial core states with the total angular
momentum $j$ to final unoccupied Bloch states
%
\begin{eqnarray}
\mu_j^{\lambda} (\omega) &=& \sum_{m_j} \sum_{n \bf k} | \langle \Psi_{n \bf k} |
\Pi _{\lambda} | \Psi_{jm_j} \rangle |^2 \delta (E _{n \bf k} - E_{jm_j} -
\hbar \omega ) 
 \theta (E _{n \bf k} - E_{\rm F} ),
\label{mu}
\end{eqnarray}
where $\Psi _{jm_j}$ and $E _{jm_j}$ are the wave function and the
energy of a core state with the projection of the total angular
momentum $m_j$; $\Psi_{n\bf k}$ and $E _{n \bf k}$ are the wave
function and the energy of a valence state in the $n$-th band with the
wave vector {\bf k}; $E_{\rm F}$ is the Fermi energy.

$\Pi _{\lambda}$ is the electron-photon interaction
operator in the dipole approximation
\begin{equation}
\Pi _{\lambda} = -e \mbox{\boldmath$\alpha $} \bf {a_{\lambda}},
\label{Pi}
\end{equation}
where $\bm{\alpha}$ are the Dirac matrices and $\bf {a_{\lambda}}$ is the
$\lambda$ polarization unit vector of the photon vector potential,
with $a_{\pm} = 1/\sqrt{2} (1, \pm \ri, 0)$,
$a_{\parallel}=(0,0,1)$. Here, ``$+$'' and ``$-$'' denotes, respectively, left-hand
and right-hand circular photon polarizations with respect to the
magnetization direction in the solid. Then, X-ray magnetic circular
and linear dichroism are given by $\mu_{+}-\mu_{-}$ and
$\mu_{\parallel}-(\mu_{+}+\mu_{-})/2$, respectively.  More detailed
expressions of the matrix elements in the electric dipole
approximation may be found in
references~\cite{GET+94,book:AHY04,AHS+04}.  The matrix elements due
to magnetic dipole and electric quadrupole corrections are presented
in reference~\cite{AHS+04}.

\subsection{General properties of spin density waves}
\label{sec_2_2}

The magnetic configuration of an incommensurate spin spiral shows the
magnetic moments of certain atomic planes varying in direction. The
variation has a well-defined period determined by a wave vector~\mbox{\boldmath$q$}. When the magnetic moment is confined to the
lattice sites, the magnetization \mbox{\boldmath$M$} varies as
\cite{EAJ+03}

\begin{equation}
\mbox{\boldmath$M$}(\mbox{\boldmath$r$}_n)=m_n \left[ 
\begin{array}{c}
\cos(\mbox{\boldmath$q r$}_n+\phi_n)\sin(\theta_n)   \\ 
\sin(\mbox{\boldmath$q r$}_n+\phi_n)\sin(\theta_n)   \\
\cos(\theta_n)  
\end{array}
\right] ,
\label{M}
\end{equation}
where the polar coordinates are used and $m_n$ is the magnetic moment
of atom $n$ with a phase $\phi_n$ at the position
$\mbox{\boldmath$r_n$}$. Here, we consider only planar spirals, that
is, $\theta_n=\piup$/2 which also give the minimum of the total
energy. The magnetization of equation~(\ref{M}) is not translationally
invariant but transforms as
%
\begin{equation}
\mbox{\boldmath$M(r+R)$}=\mbox{$D$(\boldmath$qR$)\boldmath$M(r)$} ,
\label{trans}
\end{equation}
where \mbox{\boldmath$R$} is a lattice translation and $D$ is a
rotation around the $z$ axis. A spin spiral with a magnetization in a
general point \mbox{\boldmath$r$} in space can be defined as a
magnetic configuration which transforms according to
equation~(\ref{trans}). Since the spin spiral describes a spatially
rotating magnetization, it can be correlated with a frozen magnon.

Since the spin spiral breaks the translational symmetry, the Bloch
theorem is no longer valid. Computationally, one should use large
super-cells to obtain total-energy of the spin spirals. However, when
the spin-orbit interaction is neglected, spins are decoupled from the
lattice and only the relative orientation of the magnetic moments is
important.  Then, one can define the generalized translations which
contain translations in the real space and rotations in the spin
space \cite{Sand91}. These generalized translations leave the magnetic
structure invariant and lead to a generalized Bloch theorem. Therefore,
the Bloch spinors can still be characterized by a \mbox{\boldmath$k$}
vector in the Brillouin zone, and can be written as
\begin{equation}
\psi_k(\mbox{\boldmath$r$})=\re^{\ri\mbox{\boldmath$kr$}} \left( 
\begin{array}{c}
\re^{-\ri\mbox{\boldmath$qr$}/2} u_k(\mbox{\boldmath$r$})   \\ 
\re^{+\ri\mbox{\boldmath$qr$}/2} d_k(\mbox{\boldmath$r$})
\end{array}
\right) .
\label{defM}
\end{equation}
The functions $u_k(\mbox{\boldmath$r$})$ and
$d_k(\mbox{\boldmath$r$})$ are invariant with respect to lattice
translations having the same role as for normal Bloch functions. Due
to this generalized Bloch theorem, the spin spirals can be studied
within the chemical unit cell and no large super-cells are
needed. Though the chemical unit cell can be used, the presence of
the spin spiral lowers the symmetry of the system. There remain only the space-group operations that leave invariant the wave vector of the spiral. Considering the general spin space groups, i.e.,
taking the spin rotations into account, the space-group operations
which reverse the spiral vector together with a spin rotation of $\piup$
around the $x$ axis are symmetry operations \cite{Sand91}.

Though the original formulation of the local-spin-density
approximation of the density-functional theory permited the noncollinear
magnetic order, first-principles calculations for this aspect have
begun only recently (for a review, see reference~\cite{Sand98}). One
application is the study of noncollinear ground states, for
example, in $\gamma$-Fe (references~\cite{SanKu92,BylKle99,KSK00}) or in
frustrated antiferromagnets \cite{KBH+01,HoHa00}. In addition, the
noncollinear formulation enables the studies of finite-temperature
properties of magnetic materials. Since the dominant magnetic
excitations at low temperatures are spin waves which are noncollinear
by nature, it is possible to determine the magnon spectra and
ultimately the Curie temperature from first principles~\cite{AKS+95,RoJo97,UkKu96,HEP+98,PKT+01}. The noncollinear magnetic
configurations were investigated in the Heusler alloys Ni$_2$MnGa,
Ni$_2$MnAl~\cite{EAJ+03}, IrMnAl~\cite{AHY+06}, Mn$_3$ZnC~\cite{AHY+07a} and Mn$_3$CuN~\cite{AnBe14}. The total energies for
different spin spirals were calculated and the ground-state magnetic
structures were identified.

\subsection{Crystal structure} 

\begin{figure}[tb]
\begin{center}
\includegraphics[width=0.73\columnwidth]{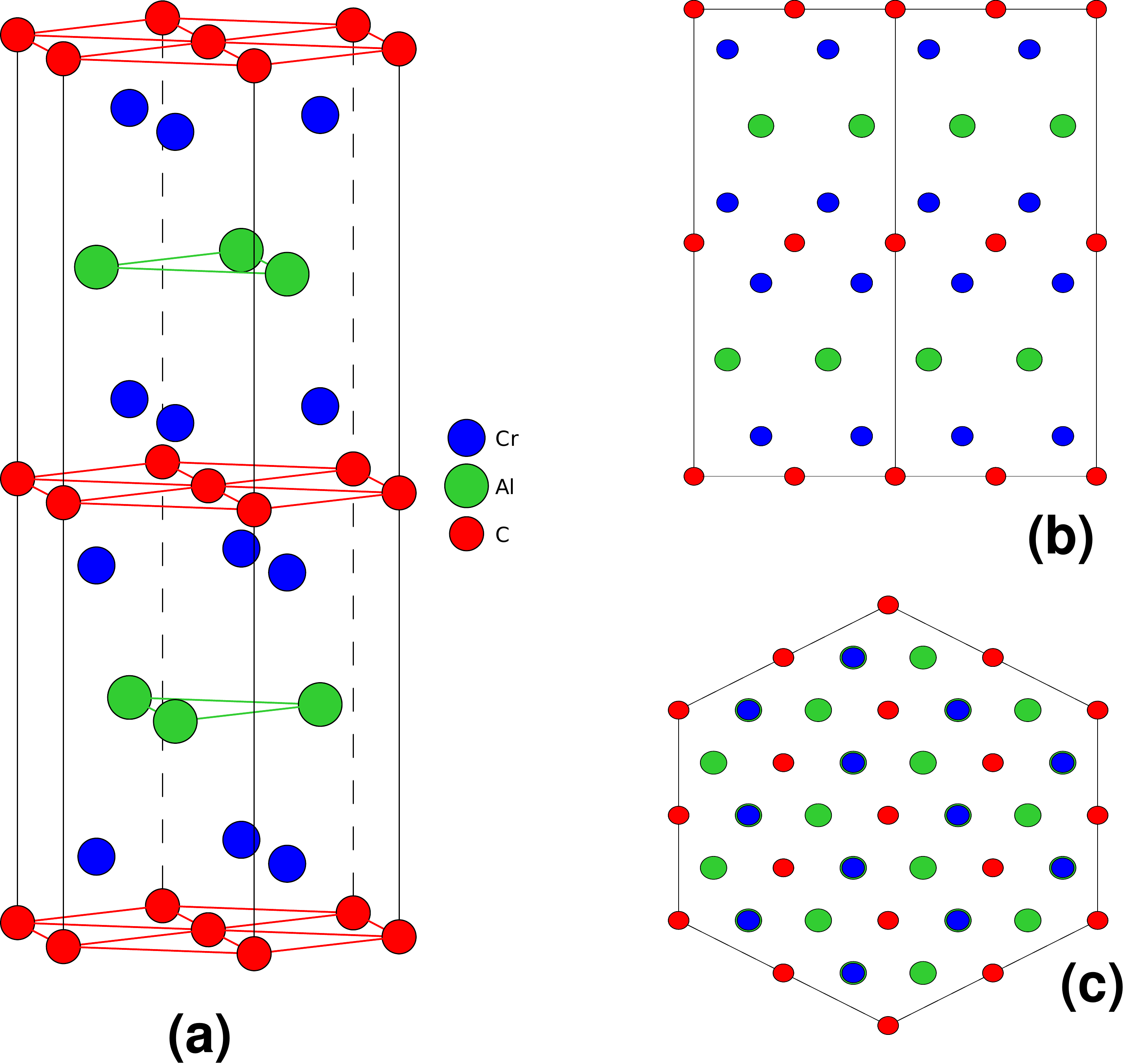}
\end{center}
\caption{\label{struc_MAX}(Colour online) (a) Crystal structure of the
  Cr$_2$AlC (the space group is $P6_3/mmc$ No. 194). Blue spheres
  represent Cr atoms, green and blue spheres show Al and C atoms,
  respectively; (b) and (c) present arrangement of atoms on a
  ($1\bar{2}10$) and (0001) planes of Cr$_2$AlC, respectively. }
\end{figure}

The M$_2$AX unit cell (the space group is $P6_3/mmc$ No. 194) with two
formula units per unit cell is shown in figure~\ref{struc_MAX} with
Wyckoff positions: M (4$f$), A (2$d$), and X (2$a$). The coordination
of the M is trigonal prismatic, while that of X is octahedral.  The
structure possesses the layer stacking sequence of M and A atoms along
the [0001] direction consisting of M$_2$X slabs and intercalation of
planar packed A-ions. The layered stacking characteristics can be
clearly illustrated in the ($1\bar{2}10$) plane, as displayed in
figure~\ref{struc_MAX}~(b). The X atoms occupy the interstitial sites of
M octahedra [see figure~\ref{struc_MAX}~(c)]. The crystal structure has
one free internal parameter by $Z_M$ that defines the height of the M
atoms above the X sheets. The dimensionless crystallographic
coordinate is $z_M = Z_M/c$. At the ideal value of $z_M =
1/12\sim0.0833$, the M and A planes are evenly spaced. Parameter $z_M
= 0.086$ and 0.083 for Cr$_2$AlC and Ti$_2$AlC, respectively.

\subsection{Calculation details}

The details of the computational method are described in our previous
papers \cite{LYA+06,AHY+07b,AYJ10,RTR+11} and here we only mention
several aspects. Band-structure calculations were performed using the
fully relativistic linear muffin-tin orbital (LMTO) method~\cite{And75,book:AHY04}. This implementation of the LMTO method uses
four-component basis functions constructed by solving the Dirac
equation inside an atomic sphere \cite{NKA+83}. The
exchange-correlation functional of a GGA-type was used in the version
of Perdew, Burke and Ernzerhof~(PBE)~\cite{PBE96}. Brillouin zone (BZ)
integration was performed using the improved tetrahedron method
\cite{BJA94}. The basis consisted of Cr and Ti $s$, $p$, $d$, and $f$
and Al and C $s$, $p$, and $d$ LMTO's.

We found that the agreement between the theoretically calculated and
experimentally measured XAS and XMCD spectra became much better with
taking into account strong Coulomb correlations. To include the
electron-electron correlations into the consideration, we used the
``relativistic'' generalization of the rotationally invariant version of
the LSDA+$U$ method \cite{YAF03} which takes into account SO coupling
so that the occupation matrix of localized electrons becomes
non-diagonal in spin indexes. This method is described in detail in
our previous paper \cite{YAF03} including the procedure to calculate
the screened Coulomb~$U$ and exchange $J$ integrals, as well as the
Slater integrals $F^2$, $F^4$, and $F^6$.

The screened Coulomb $U$ and exchange Hund coupling $J_H$ integrals
enter the LSDA+$U$ energy functional as external parameters and
should be determined independently. These parameters can be
determined from supercell LSDA calculations using Slater's transition
state technique \cite{AG91,SDA94}, from constrained LSDA calculations
(cLSDA) \cite{DBZ+84,SDA94,PEE98,CoGi05,NAY+06} or from the
constrained random-phase approximation~(cRPA) scheme~\cite{AIG+04}. Subsequently, a combined cLSDA and cRPA method was also
proposed \cite{SoIm05}. The cLSDA calculations produce $J_H = 0.95$~eV
for the Cr in Cr$_2$AlC. It is known that the cRPA method
underestimates the values of $U$ in some cases \cite{AAB+14}. On the other
hand, the cLSDA method produces too large values of $U$
\cite{AKJ+06}. Therefore, in our calculations we treated the Hubbard
$U$ as an external parameter and varied the effective Hubbard
$U_{\rm eff}= U - J_H$ from 0 to 3.0~eV. In the case of
$U_{\text{eff}}=U-J=0$, the effect of the GGA+$U$ comes from
non-spherical terms which are determined by $F^{2}$ and $F^{4}$ Slater integrals. This approach is similar to the orbital polarization (OP)
corrections \cite{Bro85,EBJ90,SBJ93,MSK98,book:AHY04}. Therefore, we
use the notation GGA+OP throughout the paper for the
$U_{\text{eff}}=U-J=0$~eV approach.

The X-ray absorption and dichroism spectra were calculated taking into
account the exchange splitting of the core levels.
The finite lifetime of a core hole was taken into account by folding the
spectra with a Lorentzian. The widths of core levels
$\Gamma_{L_{2,3}}$ for Mn and Ti, and $\Gamma_K$ for O were taken from
reference~\cite{CaPa01}. The finite experimental resolution of the
spectrometer was taken into account by a Gaussian of width~0.6~eV.

\section{Electronic and magnetic structures}
\label{sec_3}

The theoretically calculated electronic band structure of nanolaminated
ternary carbides such as Cr$_2$AlC and Ti$_2$AlC, demonstrated that
the valence bands could be divided into several regions. The lowest
lying group of valence states originates predominantly from the C 2$s$
states. The states located at higher energy range are hybridization states
of Al 3$s-3p$ orbitals. Strong $pd$ covalent bonding states derived
from Ti-C interactions dominate the adjacent higher energy range. The
states just below the Fermi level contain a relatively weaker $pd$
covalent bonding between Ti and Al. Moreover, the states near and above
the Fermi level are attributed to metal-to-metal $dd$ interactions and
antibonding states.

\begin{figure}[h]
\begin{center}
\includegraphics[width=0.55\columnwidth]{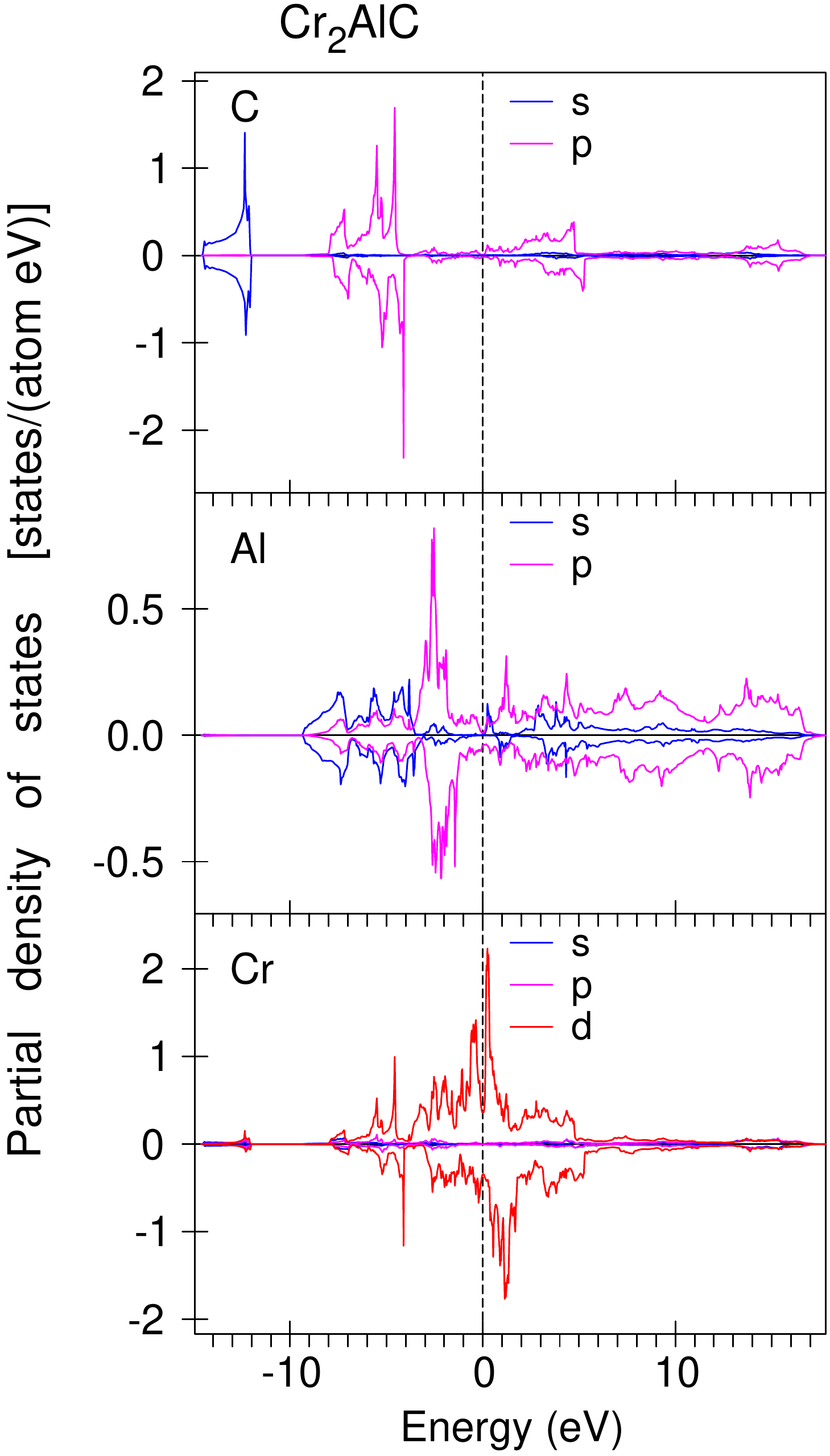}
\end{center}
\caption{\label{PDOS_CAC}(Colour online) Partial density of states [in
    states/(atom eV)] of Cr$_2$AlC in the GGA approximation. }
\end{figure}

\begin{figure}[h]
\begin{center}
\includegraphics[width=0.53\columnwidth]{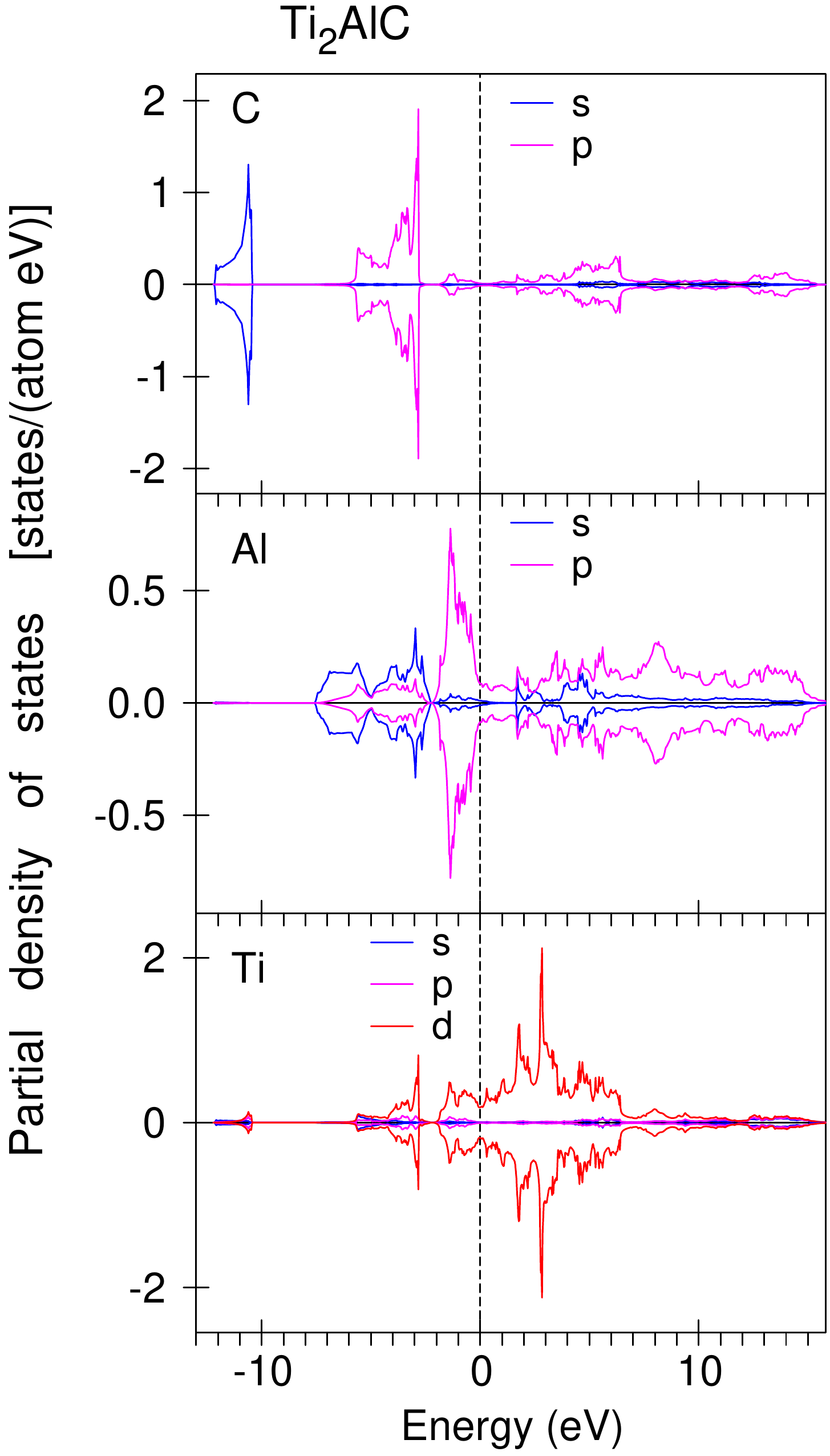}
\end{center}
\caption{\label{PDOS_TAC}(Colour online) Partial density of states [in
    states/(atom eV)] of Ti$_2$AlC in the GGA approximation. }
\end{figure}

Characteristics of atomic bonding can be vividly illustrated by
projected density of states (PDOS). Figures~\ref{PDOS_CAC} and~\ref{PDOS_TAC} show the PDOS of Cr$_2$AlC and Ti$_2$AlC for
comparison. The C 2$s$ states are located between $-14.7$~eV to
$-12.1$~eV and $-12.2$~eV to $-10.4$~eV below the Fermi level in
Cr$_2$AlC and Ti$_2$AlC, respectively. The states, which are
approximately located between $-5.8$ and $-2.8$~eV below the Fermi level~($E_{\rm F}$) in Ti$_2$AlC are C 2$p$ states. They are well hybridized with
Al $sp$ and Ti 3$d$ states. The corresponding C 2$p$ states Cr$_2$AlC
are situated lower in energy at the $-8.0$ and $-4.3$~eV energy interval.
Al 3$p$ states occupy a wide energy interval from $-9.0$~eV to $16.0$~eV and
from $-7.7$~eV to 16.0~eV Cr$_2$AlC and Ti$_2$AlC, respectively. The
major peak of the occupied Al 3$p$ states associated with the $pd$
covalent bond is situated at around $-2.5$~eV in Cr$_2$AlC and $-1.0$~eV
in Ti$_2$AlC below $E_{\rm F}$. The 3$d$ orbitals of Cr atoms dominate the
states near the $E_{\rm F}$, and the contribution from Al $p$-derived
orbitals is negligible. In comparison, the states near $E_{\rm F}$ are
dominated by Ti 3$d$ orbitals in Ti$_2$AlC, with some contribution
from Al $p$ orbitals.

The magnitude of Cr spin and orbital magnetic moments strongly
depends on the Hubbard $U$. The GGA approach produces the spin and
orbital moments equal to 0.707~{\mb} and 0.007~{\mb}, respectively
(see table~\ref{mom_MAX}). The GGA+OP approach gives slightly larger
moments: $M_s^{\rm Cr}=0.757$~{\mb} and $M_l^{\rm Cr}=0.009$~{\mb}.  For
$U_{\rm eff} = 1$~eV, 2~eV and 3~eV, the spin magnetic moments are equal
to 0.973~{\mb}, 2.097~{\mb}, and 2.799~{\mb}, respectively.  These
values are in good agreement with previous band structure calculations~\cite{RLA11,DAR13,DAR15}. The orbital magnetic moments at the Cr site also strongly increased with increasing the value of Hubbard~$U$
and even change the sign for the $U_{\rm eff}$ = 2~eV and 3~eV (table
\ref{mom_MAX}). We found a similar dependence of the spin and
orbital magnetic moments for the AFM solution as well, although the
absolute values of the moments are slightly different from the FM
ordering. The induced spin magnetic moments at the Al and~C sites are much
smaller than at the Cr site and have an opposite direction (table
\ref{mom_MAX}). It is interesting to note that the spin magnetic moment at
the C site is larger than the corresponding moment at the Al
site. Orbital moments on both the Al and C sites are quite small.

\begin{table}[tbp!]
\caption{\label{mom_MAX}The theoretically calculated spin $M_s$,
  orbital $M_l$, and total magnetic moments (in~{\mb}) of Cr$_2$AlC
  for the FM solution.}
\begin{center}
\begin{tabular}{ccccccccc}
\hline
Method       & Atom & $M_s$  & $M_l$ & $M_{\rm total}$ \\
\hline
              &  Cr   &  0.707 &  0.007  &  4.134  \\
GGA           &  Al   & $-0.015$ & $-0.001$  & $-0.016$  \\
              &   C   & $-0.064$ &  0.0    & $-0.064$  \\
\hline
              &  Cr   &  0.757 &  0.009  &  0.766  \\
GGA+OP        &  Al   & $-0.021$ & $-0.001$  & $-0.022$  \\
              &   C   & $-0.064$ &  0.0    & $-0.064$  \\
\hline
              &  Cr   &  0.973 &  0.013  &  0.986  \\
GGA+$U$       &  Al   & $-0.074$ & $-0.001$  & $-0.075$  \\
$U_{\rm eff}=1$ eV &   C   & $-0.108$ &  0.0    & $-0.108$  \\
\hline
              &  Cr   &  2.097 & $-0.030$  &  2.067  \\
GGA+$U$       &  Al   & $-0.154$ & $-0.001$  & $-0.155$ \\
$U_{\rm eff}=2$ eV &   C   & $-0.239$ & $-0.003$  & $-0.242$  \\
\hline
              &  Cr   &  2.799 & $-0.030$  &  2.769  \\
GGA+$U$       &  Al   & $-0.196$ & $-0.001$  & $-0.197$  \\
$U_{\rm eff}=3$ eV &   C   & $-0.357$ & $-0.002$  & $-0.359$  \\
\hline
LSDA reference~\cite{SSM+04} &  Cr  & 0.70 & $-$  & $-$  \\
\hline
GGA+$U$ reference~\cite{RLA11} &  Cr  & 0.90 &$-$  & $-$  \\
$U_{\rm eff}=1$ eV &      &  &   &   \\
\hline
GGA+$U$ reference~\cite{RLA11} &  Cr  & 2.50 & $-$ & $-$  \\
$U_{\rm eff}=2$ eV &      &  &   &   \\
\hline
exper.$_{\rm{XMCD}}$ \cite{JBJ+14} &  Cr   &  0.05 & $-$ &  $-$  \\
\hline
exper.$_{\rm{SQIID}}$ \cite{JCC+13} &  Cr   &  0.002 & $-$ &  $-$  \\
\hline
\end{tabular}
\end{center}
\end{table}

SQUID experiments by Jaouen et~al. \cite{JCC+13} produce
Cr$_2$AlC to be FM with extremely small Cr spin magnetic moment of
0.002~{\mb}. Recent X-ray magnetic circular dichroism experiments
\cite{JBJ+14} clearly demonstrate that Cr atoms carry a net magnetic
moment of 0.05~{\mb} in Cr$_2$AlC along $c$ axis. Therefore, these
experiments are in strong disagreement with all existing theoretical
calculations concerning the value of the Cr spin magnetic moment and
the exact nature of the magnetic ordering in Cr$_2$AlC ternary phase.

In order to gain an insight into the ground state magnetic configurations
of Cr$_2$AlC, we compared the calculated total energies for different
spin ordering Cr$_2$AlC, namely, NM, FM, and AFM phases as well as
possible noncollinear magnetic (NCM) structures. We found that the
total energy of the AFM$_{ab}$ state (AFM ordering in the $ab$ plane)
is lower by 25~meV/f.u., 17~meV/f.u., and 12~meV/f.u. than those of
the NM, FM, and AFM$_{[0001]}$ (AFM ordering along the $c$ axis)
states, respectively, which implies a preferable AFM phase in
Cr$_2$AlC ordered in the $ab$ plane. This conclusion agrees with the
previous estimations by Dahlqvist et~al.  \cite{DAR13,DAR15}.
However, this AFM$_{ab}$ ground state is still in contradiction with
the XMCD experiment by Jaouen et~al. \cite{JBJ+14}, because any
AFM ordering produces zero net magnetization in the system. Applying
the noncollinear formalism described in section~\ref{sec_2_2} to the Cr$_2$AlC,
we found NCM state which has a lower total energy by 6~meV/f.u. in
comparison with the AFM$_{ab}$ state. The NCM state is characterized
by a canted AFM spin configuration (spin magnetic moment $M_S^{\rm Cr}=
0.768$~{\mb} in the GGA approach with the polar angles equal to
$\theta^{\rm Cr_1}=91.7^{\circ}$, $\phi^{\rm Cr_1}=0^{\circ}$ and
$\theta^{\rm Cr_2}=91.4^{\circ}$, $\phi^{\rm Cr_1}=180^{\circ}$). Such a
AFM configuration with the spins slightly canted out of the ($a,b$) plane
produces a small projection of the Cr spin magnetic moment along the
$c$ axis of around 0.047~{\mb}, which is in excellent agreement with
the estimation by Jaouen et~al.  \cite{JBJ+14} of 0.05~{\mb}
using the XMCD measurements and sum rules. Strictly speaking, we 
do not have AFM ordering in the ($a,b$) plane but rather ferrimagnetic (FiM)
ordering with a small net magnetization of 0.005~{\mb} in the plane.

We found that Ti$_2$AlC is very close to NM ground state. Though
titanium is in the Ti$^{2+}$ state ($d^2$), the spin magnetic moment at
the Ti as well as at Al sites is less than 10$^{-4}$~{\mb}.

\section{X-ray absorption and XMCD spectra}
\label{sec_4}

\begin{figure}[h!]
	\begin{center}
		\includegraphics[width=0.50\columnwidth]{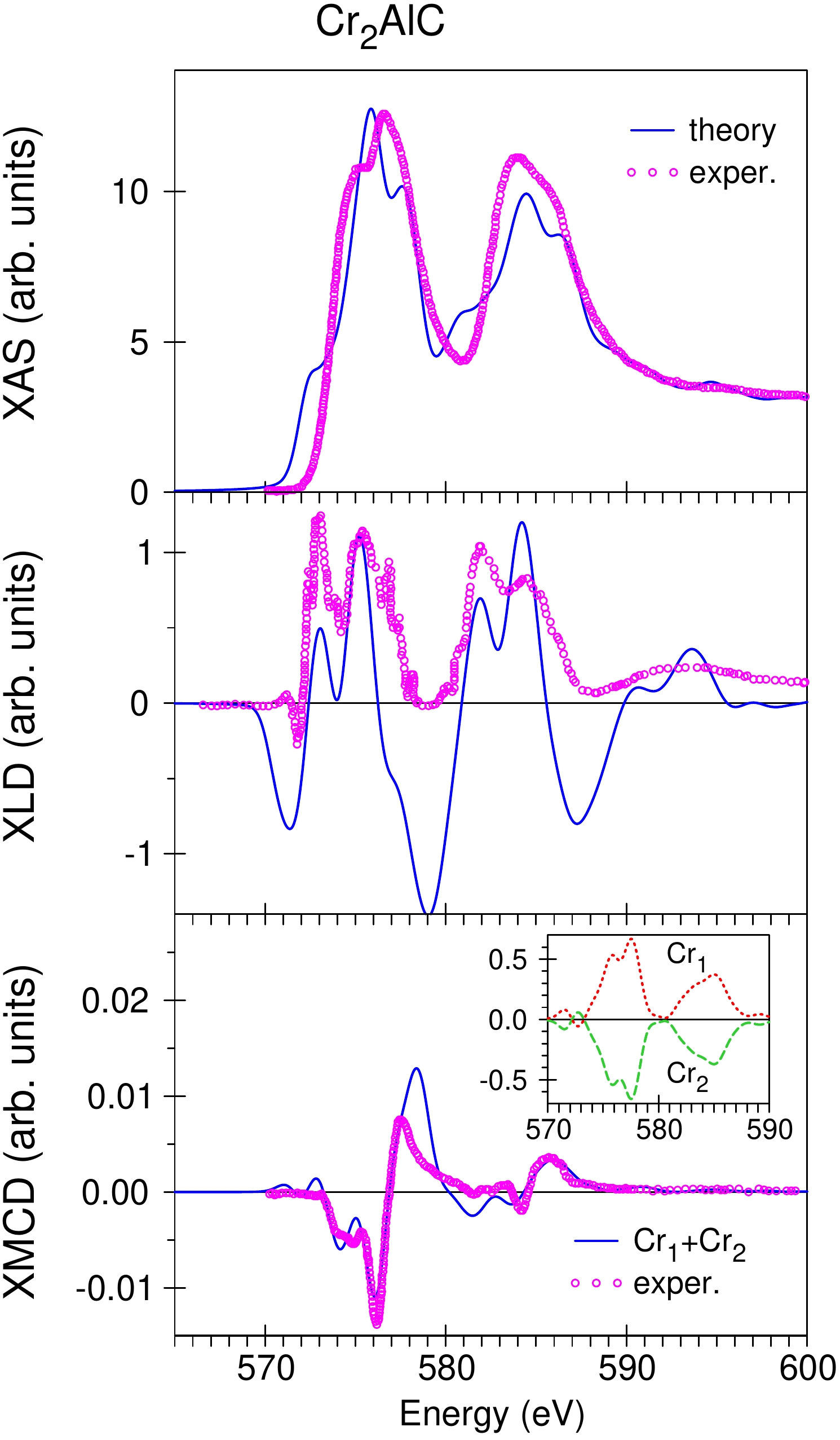}
	\end{center}
	\caption{\label{Cr_L23_CAC}(Colour online) Top panel: the X-ray
		absorption spectra (open circles) at the Cr $L_{2,3}$ edges in
		Cr$_2$AlC measured at 4.2 K \cite{JBJ+14} with a 6 T magnetic
		compared with the theoretically calculated ones (full blue curve)
		calculated in the GGA approach; Middle panel: the theoretically
		calculated (full blue curve) and experimentally measured (open
		circles) \cite{JBJ+14} X-ray linear dichroism spectra; lower panel:
		the XMCD experimental spectra (open circles) of Cr$_2$AlC at the Cr
		$L_{2,3}$ edges and the theoretically calculated one (full blue
		line); the onset shows the XMCD spectra at the Gr $L_{2,3}$ edges
		separately from the Cr$_1$ and Cr$_2$ sites. }
\end{figure}

Figure \ref{Cr_L23_CAC} presents the X-ray absorption spectra (open
circles) at the Cr $L_{2,3}$ edges (top panel) in Cr$_2$AlC measured
at 4.2~K \cite{JBJ+14} with a 6~T exteranal magnetic field compared
with the theoretically calculated ones (full blue curve). Since the
pure Cr $L_{23}$ edges are structureless, the existence of these fine
structures shows that the chromium is not in a pure metallic state in
Cr$_2$AlC, a well-known fact from previous band structure calculations
related to MAX phases \cite{LZL07}. They all have a mixture of
covalent, metallic and ionic bonds.

One can observe a well separated double structure at the top of both
$L_3$ and $L_2$ edges. These fine structures are well reproduced by a 
band structure calculation, although the theory produces inverse
relative intensities between low and high energy peaks in comparison
with the experimentally observed ones in the $L_3$ XAS spectrum. It is
interesting to note that the model multiple scattering calculations
performed with the FEFF \cite{AKG+09,ReAl00} code by Jaouen et~al.~\cite{JBJ+14} show the same inversion in the relative
intensities as obtained by us.

Figure \ref{Cr_L23_CAC} (middle panel) shows the theoretically
calculated (full blue curve) and experimentally measured (open
circles) \cite{JBJ+14} X-ray linear dichroism spectra. The theory
reproduces the energy position of all the fine structures quite well, although the negative peaks at 571~eV, 578~eV, and 587~eV are much
lower in theory than in the experiment.

The XMCD experimental spectra (open circles) of Cr$_2$AlC at the Cr
$L_{2,3}$ edges and the theoretically calculated spectrum (full blue line)
are presented in the lower panel of figure~\ref{Cr_L23_CAC}. For the
experimental geometry used in reference~\cite{JBJ+14} it follows that the
electric field, {\bf E}, of the incident X-ray beam is parallel to the
$(a,b)$ plane of the MAX phase. In that case, XAS measurements mainly
probe the unoccupied Cr 3{\dxy}, 3{\dxz}, 3{\dyz} and 3{\dxxyy} in-plane
orbitals, as well as a much smaller $p \to s$ contribution. The X-ray
dichroism at the Cr $L_{2,3}$ edges is quite small due to cancellation
signals from Cr$_1$ and Cr$_2$ sites (see the insert in the lower panel of
figure~\ref{Cr_L23_CAC}). The agreement between the theory and
experiment is quite good although the intensity of the major positive peak at 578
eV is slightly overestimated in the theory.

\begin{figure}[h!]
	\begin{center}
		\includegraphics[width=0.60\columnwidth]{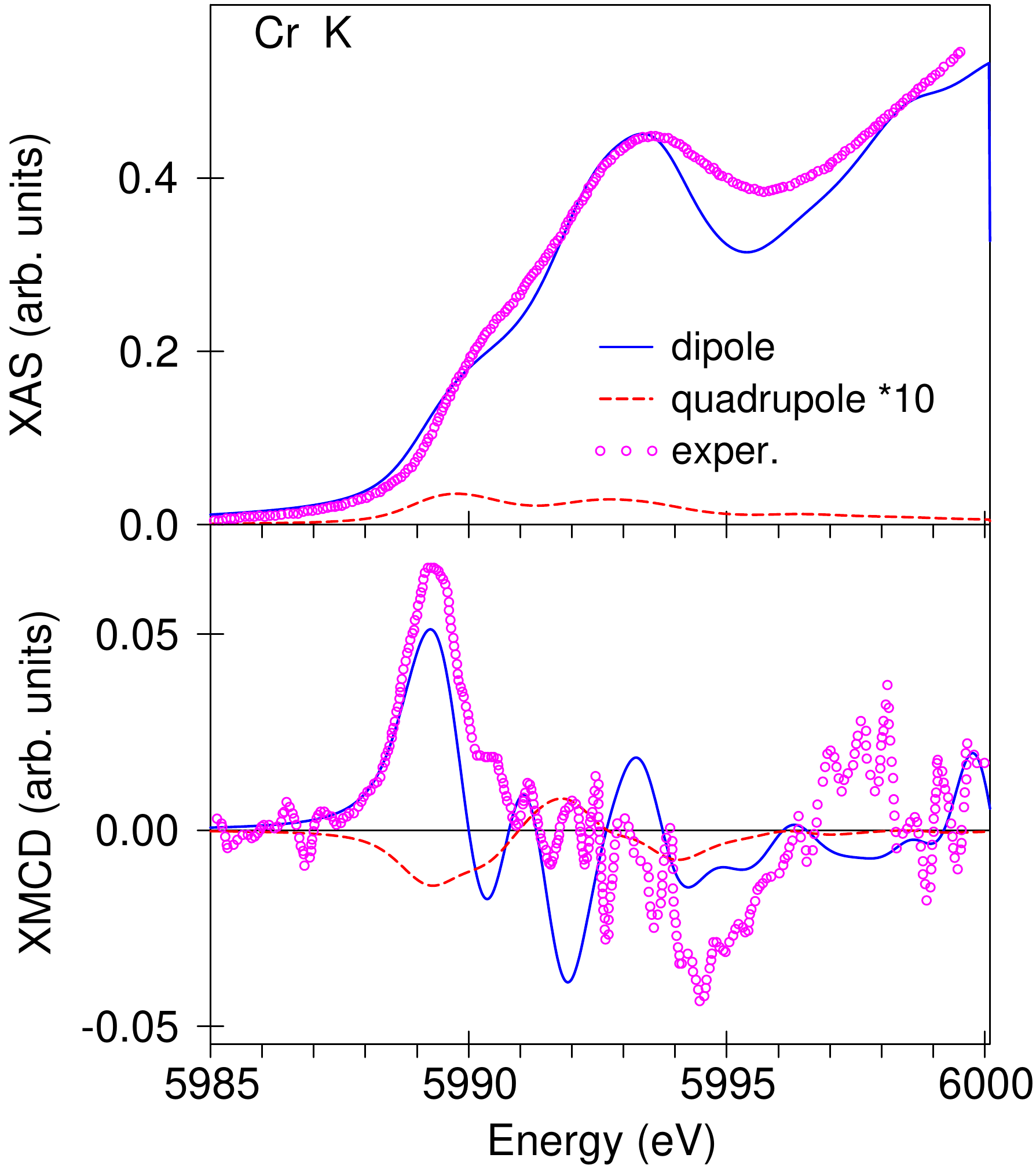}
	\end{center}
	\caption{\label{Cr_K_CAC}(Colour online) Top panel: the X-ray
		absorption spectrum (open circles) at the Cr $K$ edge in Cr$_2$AlC
		measured at 2.2~K \cite{JBJ+14} and external magnetic field of 10~T
		with the theoretically calculated ones in the GGA approach taking
		into account only dipole 1$s$ $\to$ 2$p$ transitions (full blue
		curve) and quadrupole 1$s$ $\to$ 3$d$ transitions (dashed red curve)
		multiply by factor 10; lower panel: the XMCD experimental spectrum
		(open circles) of Cr$_2$AlC at the Cr $K$ edge and the theoretically
		calculated ones in dipole approximation (full blue line) and
		quadrupole transitions (dashed red curve) multiply by factor~10. }
\end{figure}

We found that both the GGA and GGA+$U$ with $U_{\rm eff} \leqslant 1$~eV give the 
calculated XAS and XMCD spectra close to experimentally measured
ones. For larger values of the $U$-parameter ($U_{\rm eff} >1$~eV), the
theoretically calculated XMCD spectra deviate strongly from
experimentally observed spectra. Therefore, this class of Cr-based
carbide MAX phases cannot be considered as strongly correlated
systems. A similar conclusion was also drawn by Dahlqvist et~al.
\cite{DAR13,DAR15} based on the calculations of lattice parameters
and bulk modulus in the Cr$_2$AlC, Cr$_2$GaC and Cr$_2$GeC.

Figure~\ref{Cr_K_CAC} presents the X-ray absorption spectrum (open
circles) at the Cr $K$ edge (top panel) in Cr$_2$AlC measured at 2.2~K
\cite{JBJ+14} with magnetic field of 10~T compared with the
theoretically calculated ones (full blue curve). The XAS spectrum
consists of  a major peak at 5993.4~eV followed by a local minimum at 5996.0~eV and a low energy shoulder at 5989.3~eV. The theory well reproduces
the energy position and shape of the fine structures. It is worth
noticing that the low energy shoulder at 5989.3~eV is usually
attributed to the quadrupolar $E_2$ (1$s \to 3d$) transitions
\cite{ReAl00,JBJ+14}.  We investigate the effect of the electric
quadrupole $E_2$ and magnetic dipole $M_1$ transitions. We found that
the $M_1$ transitions are extremely small in comparison with the $E_2$
transitions and can be neglected. The $E_2$ transitions indeed
contribute to the low energy shoulder at 5989.3~eV as well as to the
major peak at 5993.4~eV, although the quadrupolar~$E_2$ transitions
are by two orders of magnitude smaller than the electric dipole
transitions $E_1$ (see a red dashed curve at the upper panel of
figure~\ref{Cr_K_CAC}). Therefore, the low energy shoulder reflects the
energy distribution of C $N_p$ partial DOS (figure~\ref{PDOS_CAC}).

The lower panel shows the XMCD experimental spectrum (open circles) of
Cr$_2$AlC at the Cr $K$ edge and the theoretically calculated ones in the dipole approximation (full blue line). The dashed-dotted lines show
the contribution from the quadrupole $E_2$ (1$s \to 3d$) transitions
multiplied by a factor of 10. Again, the contribution of the
quadrupolar $E_2$ (1$s \to 3d$) transitions to the Cr $K$ XMCD
spectrum is very small. The theory well reproduces the energy position
and the intensity of the major positive peak at 5989.3~eV, while other fine
structures are reproduced with less accuracy, although it is hard to
achieve an ideal agreement with the experimental measurements with such a
very weak detected XMCD signal.

\begin{figure}[h!]
\begin{center}
\includegraphics[width=0.60\columnwidth]{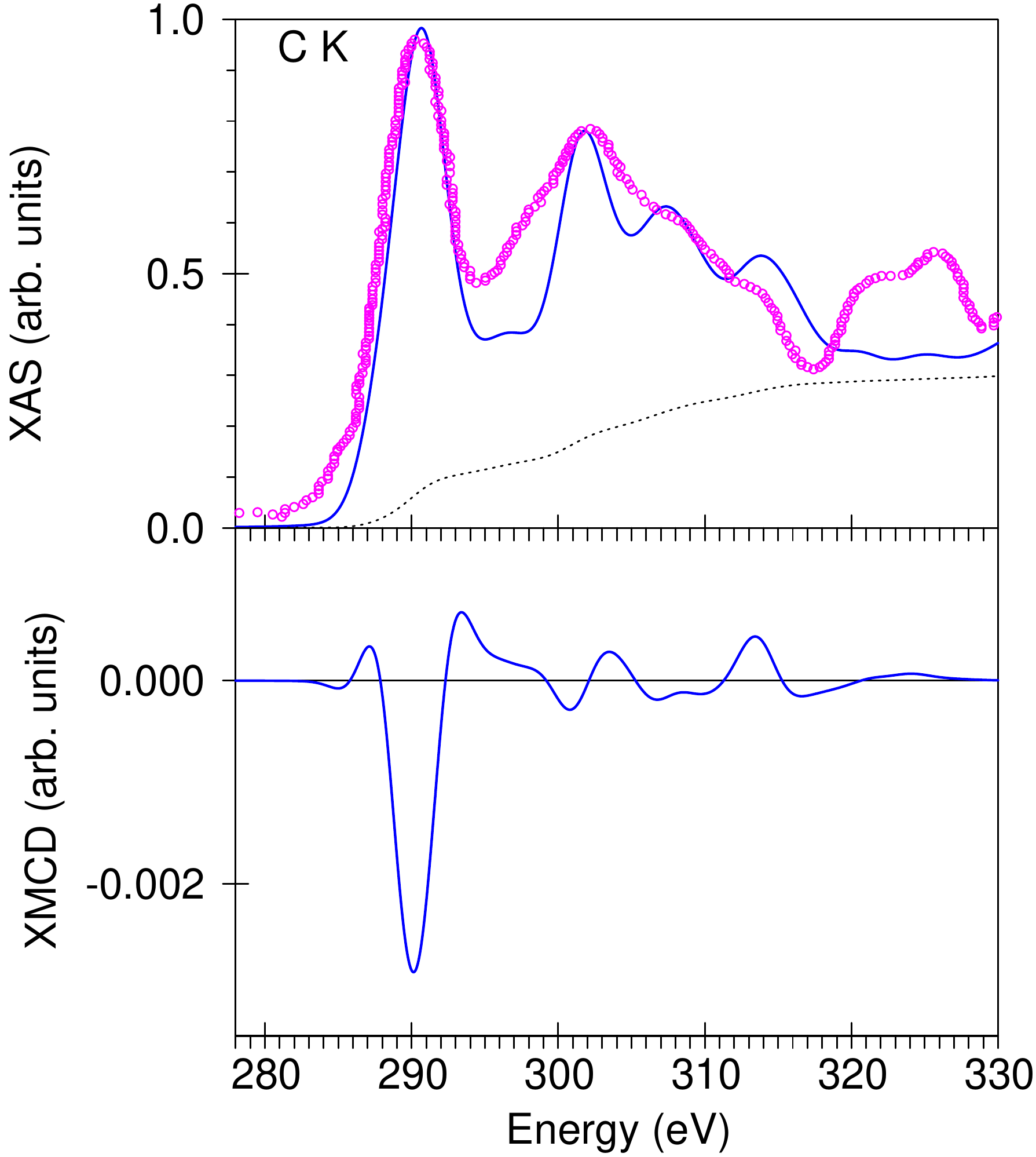}
\end{center}
\caption{\label{C_K_CAC}(Colour online) Top panel: the X-ray absorption
  spectrum (open circles) at the C $K$ edge in Cr$_2$AlC measured at
  4.2~K \cite{LZZ+06} with a 6 T magnetic compared with the
  theoretically calculated one (full blue curve) in the GGA approach;
  lower panel: the theoretically calculated XMCD spectrum of Cr$_2$AlC
  at the C $K$ edge. }
\end{figure}

Figure \ref{C_K_CAC} presents the X-ray absorption spectrum (open
circles) at the C $K$ edge (top panel) in Cr$_2$AlC measured by Lin et~al.~\cite{LZZ+06} compared with the theoretically calculated
ones (full blue curve). Our band structure calculations well reproduce
the energy position of all fine structures of the experimental C~$K$~XAS spectrum. Shindo and Oikawa \cite{book:ShOi99} investigated the
XAS spectra of diamond, graphite, and amorphous carbon. They reported
that a XAS peak at 291~eV indicated a strong $\sigma$ bonding state
for C. On the other hand, a $\pi$ bonding state of carbon was
distinguished by a $\pi^*$ peak at $\sim284$~eV in the XAS spectra
\cite{LZZ+06,book:ShOi99}. In the experimentally measured C $K$ edge
for Cr$_2$AlC, the peak was located at about 290~eV, which implies
that the Cr--C bond in Cr$_2$AlC is a strong $\sigma$ bonding
\cite{LZZ+06}.

Figure \ref{C_K_CAC} (bottom panel) presents the theoretically
calculated XMCD spectrum at the C $K$ edge in Cr$_2$AlC. Due to very
small spin and orbital magnetic moments at the C site (see table~\ref{mom_MAX}), one would expect a quite small dichroism at this edge
with major negative peak at 291~eV. The experimental measurements of
the XMCD spectrum at the C $K$ edge are highly desirable.
 
Spin and orbital magnetic moments in Ti$_2$AlC are very small at the
Ti and C sites. Therefore, the XMCD spectra at these edges are not
detected yet. The theoretically calculated XMCD spectra (not shown)
are three orders of magnitude smaller than their XAS spectra.

\begin{figure}[h!]
	\begin{center}
		\includegraphics[width=0.60\columnwidth]{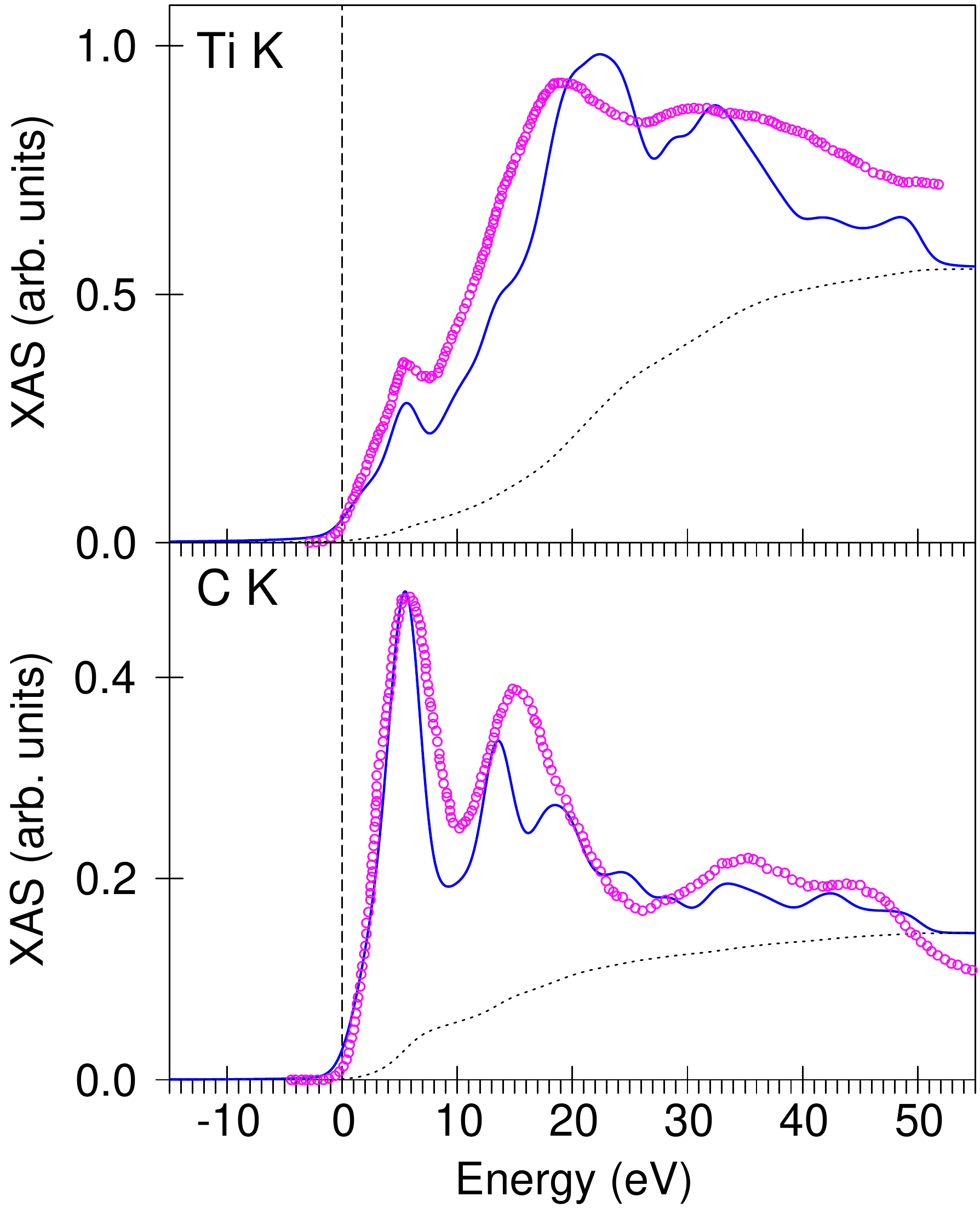}
	\end{center}
	\caption{\label{Ti_C_K_TAC}(Colour online) top panel: the
		experimentally measured \cite{HJB05} X-ray absorption spectrum at Ti
		$K$ edge in Ti$_2$AlC (magenta circles) and theoretically calculated
		(full blue curve) in the GGA approach; lower panel: the
		experimentally measured \cite{HJB05} X-ray absorption spectrum at C
		$K$ edge in Ti$_2$AlC (magenta circles) and theoretically calculated
		(full blue curve) in the GGA approach. }
\end{figure}

Figure \ref{Ti_C_K_TAC} shows the X-ray absorption spectra (open
circles) at the Ti $K$ (upper panel) \cite{HJB05} and C $K$ (lower
panel) \cite{HJB05} edges in Ti$_2$AlC compared to the
theoretically calculated spectra (full blue curves). The theory quite
well reproduces the experimental spectra.

The energy position of a major low energy peak at the C $K$ edge 
coincides with the low energy shoulder of Ti $K$ XAS spectrum at 4~eV
above the edge indicating a strong Ti $d$ -- C $P$ $\sigma$ bonding in
Ti$_2$AlC.

\newpage

\section{Summary}
\label{sec:summ}

The electronic and magnetic structures and X-ray magnetic circular
dichroism of the MAX 211 compounds Cr$_2$AlC and Ti$_2$AlC were
investigated theoretically within GGA and GGA+$U$ approaches in the
framework of the fully relativistic spin-polarized Dirac LMTO
band-structure method.

We found the non-collienear magnetic state as a ground state in
Cr$_2$AlC which is characterized by a canted AFM spin configuration
(spin magnetic moment $M_S^{\rm Cr}= 0.768$~{\mb} in the GGA approach with
the polar angles equal to $\theta^{\rm Cr_1}=91.7^{\circ}$,
$\phi^{\rm Cr_1}=0^{\circ}$ and $\theta^{\rm Cr_2}=91.4^{\circ}$,
$\phi^{\rm Cr_1}=180^{\circ}$). Such AFM configuration with spins
is slightly canted out of the ($a,b$) plane and produces a small
projection of the Cr spin magnetic moment along the $c$ axis of around
0.047~{\mb}, which is in excellent agreement with the estimation by
Jaouen et~al.  \cite{JBJ+14} of 0.05~{\mb} using the XMCD
measurements and sum rules. There is a ferrimagnetic ordering in the
($a,b$) plane with net magnetization of 0.005~{\mb}.

We have studied an X-ray magnetic circular dichroism at the Cr $L_{2,3}$
and Cr, C, and Ti $K$ edges in Cr$_2$AlC and Ti$_2$AlC. The
calculations show a good agreement with the experimental
measurements. We cannot validate the significance for using the
LSDA(GGA)+$U$ methods for the study of magnetic MAX phases since both
GGA and GGA+$U$ with $U_{\rm eff} \leqslant 1$~eV give the calculated XAS and XMCD
spectra close to the experimentally measured ones. Therefore, this class
of Cr-based carbide MAX phases cannot be considered as strongly
correlated systems.



 

\newcommand{\noopsort}[1]{} \newcommand{\printfirst}[2]{#1}
  \newcommand{\singleletter}[1]{#1} \newcommand{\switchargs}[2]{#2#1}

\newpage

	\ukrainianpart

\title
{Електронна структура та рентгенівський магнітний циркулярний дихроїзм у МАХ фазах T$_2$AlC (T$=$Ti або Cr), визначені з перших принципів
}
\author{Л. В. Бекеньов, С. В. Мокляк, Б. Ф. Журавльов, Ю. Кучеренко, В. М. Антонов}

\address{Інститут металофізики ім. Г. В. Курдюмова НАН України, бульвар Академіка Вернадського, 36, UA-03142 Київ, Україна  
}
\makeukrtitle

\begin{abstract}
	Ми вивчаємо електронні та магнітні властивості сполук T$_2$AlC (T$=$Ti та Cr) в рамках теорії функціоналу густини з використанням узагальненого градієнтного наближення (GGA) з урахуванням сильних кулонівських кореляцій (GGA+U), розраховуючи зонну структуру в формалізмі повністю релятивістського спін-поляризованого методу діраківських лінійних МТ-орбіталей (LMTO). Теоретично досліджені рентгенівські спектри поглинання та рентгенівський магнітний циркулярний дихроїзм (XMCD) на Cr $L_{2,3}$ та Cr, Ti, C $K$ краях поглинання. Результати розрахунків добре узгоджуються з експериментальними даними. Дослід\-жено вплив електричного квадрупольного $E_2$ та магнітного дипольного $M_1$ переходів на Cr $K$ краю пог\-линання.

	\keywords 
	електронна структура, спектри рентгенівського поглинання, рентгенівський магнітний циркулярний дихроїзм, МАХ фази
\end{abstract}

\lastpage
\end{document}